\documentclass[epsfig,graphics,usenatbib]{mn2e}

\usepackage{graphicx}
\usepackage{times}
\usepackage{psfrag}
\usepackage{amsmath,amssymb}
\usepackage{threeparttable}

\begin{document}

\newcommand{\Zsolar}{\mbox{\,$\rm Z_{\odot}$}}
\newcommand{\Msolar}{\mbox{\,$\rm M_{\odot}$}}
\newcommand{\Lsolar}{\mbox{\,$\rm L_{\odot}$}}
\newcommand{\ang}{\mbox{$\rm \AA$}}
\newcommand{\xs}{$\chi^{2}$}
\newcommand{\ls}{{\tiny \( \stackrel{<}{\sim}\)}}
\newcommand{\gs}{{\tiny \( \stackrel{>}{\sim}\)}}
\newcommand{\asec}{$^{\prime\prime}$}
\newcommand{\amin}{$^{\prime}$}
\newcommand{\lx}{L$_{X}$}
\newcommand{\tx}{T$_{X}$}
\newcommand{\ea}{E$+$A}
\newcommand{\ha}{H$\alpha$}
\newcommand{\oii}{[OII]}

\title[YSPs in ETGs in the SDSS]{Young stellar populations in early-type galaxies in the Sloan Digital Sky Survey}
\author[L.A. Nolan, S. Raychaudhury \& A. Kab\'{a}n]
{Louisa A. Nolan$^{1}$, Somak Raychaudhury$^{1}$ \& Ata Kab\'{a}n$^{2}$ \\
$^{1}$School of Physics and Astronomy, University of Birmingham, Birmingham, B15 2TT, UK\\
$^{2}$School of Computer Science, University of Birmingham, Birmingham, B15 2TT, UK}

\date{Submitted for publication in MNRAS}

\maketitle
  
\begin{abstract}

We use a purely data-driven rectified factor analysis to identify
early-type galaxies with recent star formation in Data Release 4 of
the Sloan Digital Sky Survey Spectroscopic Catalogue. We compare the
spectra and environment of these galaxies with those of `normal' early-type
galaxies, and a sample of independently selected \ea\ galaxies. We
calculate the projected local galaxy surface density from the nearest
five and ten neighbours ($\Sigma_{5}$ and $\Sigma_{10}$) for each
galaxy in our sample, and find that the dependence 
on projected local density, of the properties
of \ea\ galaxies, 
is not significantly different
from that of 
 early-type galaxies with young stellar
populations, 
dropping off rapidly towards denser environments, and flattening off
at densities \ls\ 0.1$-$0.3 Mpc$^{-2}$. The dearth of \ea\ galaxies in
dense environments confirms that \ea\ galaxies are most likely the
products of galaxy-galaxy merging or interactions, rather than
star-forming galaxies whose star formation has been quenched by
processes unique to dense environments, such as ram-pressure stripping
or galaxy harassment. We see a tentative peak in the number of \ea\
galaxies at $\Sigma_{10} \sim 0.1-0.3$ Mpc$^{-2}$, which may represent
the local galaxy density at which the rate 
of galaxy-galaxy merging or
interaction rate peaks. Analysis of the spectra of our early-type
galaxies with young stellar populations suggests that they have a
stellar component dominated by F stars, $\sim$ 1$-$4 Gyr old, together
with a mature, metal-rich population characteristic of `typical'
early-type galaxies. The young stars represent \gs 10\% of the stellar
mass in these galaxies. This, together with the similarity of the
environments in which this `E$+$F' population and the \ea\ galaxy
sample are found, suggests that E+F galaxies used to be \ea\ galaxies,
but have evolved by a further $\sim$ one to a few Gyr. Our rectified
factor analysis is sensitive enough to identify this hidden population,
which allows us to study the global and intrinsic properties of
early-type galaxies created in major mergers or interactions, and
compare them with those early-types which have had the bulk of their
stars in place since a much earlier epoch.

\end{abstract}


\section{introduction}

There is still much uncertainty about the relative epochs of mass assembly and star formation in early-type galaxies (e.g. de Lucia et
al. 2006; Cimatti, Daddi \& Renzini 2006). The most massive galaxies
contain the oldest stellar populations, but hierarchical structure
formation requires that their present-day mass was assembled at a
recent epoch. On the other hand, galaxy `downsizing' (e.g. Cowie et
al. 1996; Treu et al. 2005) means that less massive galaxies have more
recent star formation.

One would expect that different formation histories would give rise to different observed characteristics in early-type galaxies. Several studies (e.g. Faber et al. 1997; Emsellem et al. 2004; Khochfar \& Burkert 2005) have in fact found that various sub-classes of early-type galaxy exist (e.g. `core' and `cuspy' galaxies, fast and slow rotators). We would like to know how these sub-classes are related to the various modes of formation of early-type galaxies, and to the effects of their immediate environments.

The stellar populations of early-type galaxies hold the key to
understanding the difference between early-types created via different
evolutionary routes. If we can identify those galaxies which have
recently undergone a strong starburst, and we associate that recent
starburst with a major galaxy-galaxy merger or interaction, then we
can identify the products of
recent merger or interaction, map their evolution
with environment and redshift, and study their intrinsic properties in
comparison with more massive `primordial' early-types.

In this work, we are particularly interested in the relationship
between \ea\ galaxies and `typical' early-types. \ea\ galaxies are
defined as having strong hydrogen Balmer absorption lines, but no
corresponding [OII] or \ha\ emission lines. Their spectra, therefore, resemble 
the linear superposition of a characteristic early-type
galaxy spectrum and that of an A star. This suggests that these
galaxies have undergone a burst of star formation in the last Gyr or
so, but that there is no on-going star formation. It is hypothesised
that these galaxies will eventually settle down as `normal'
passively-evolving early-types.

\ea\ galaxies 
may therefore represent an intermediate stage between gas-rich,
disc-dominated, rotationally-supported, star-forming systems on the
one hand, and bulge-dominated, pressure-supported, passive systems on
the other. Hence, understanding the physical processes by which the
star formation is triggered and subsequently truncated, is a vital
stage in understanding the evolution of early-type galaxies.

There are three possible scenarios which could explain the existence
of \ea\ galaxies. The first is that they are in fact dust-obscured
star-forming galaxies. This can be verified by observing at radio
wavelengths, where star formation can be detected, but dust obscuration
is not significant. Miller \& Owen (2001) and Goto (2004) observed
between them a total of 51 \ea\ galaxies. Moderate star formation was
detected in only two of these galaxies, so it is unlikely that
the majority of \ea s are dusty starbursts.

More plausibly, then, \ea\ galaxies may be the product of disk-disk
merging or interaction, at a stage when the starburst has burnt
itself out, or they may be galaxies whose star formation has been
initiated and then quenched by, for example, ram-pressure stripping or
tidal stripping. The environment of these galaxies therefore holds the
key to understanding them. Although, at intermediate redshift (0.33 $<$
z $<$ 0.83), Tran et al. (2003) find that \ea s make up a
non-negligible fraction ($\sim$7--13\%) of cluster members, suggesting
a cluster-related mechanism for the truncation of star formation, such
as ram-pressure or tidal stripping, at lower redshift (z $\sim$ 0.1),
\ea s appear to occur in lower-density environments, indicative of
merging or interaction processes (Zabludoff et al. 1996; Blake et
al. 2004). In addition, both Zabludoff et al. (1996) and Blake et al. (2004) found a significant number of their \ea\ galaxies displayed tidal features and disturbed morphologies, typical of merger remnants. More recently, Goto (2005) found that \ea s selected from
the Sloan Digital Sky Survey (SDSS) Data Release~2 have an excess of
local galaxy density only at scales of $<$ 100 kpc, and not at the
scale of either clusters or larger-scale structures. This is also
highly suggestive of a merger or interaction origin for \ea s.

Conventional methods of identifying \ea\ galaxies use measurements of the equivalent width of hydrogen Balmer absorption lines (typically \ha\ and H$\delta$) and [OII] emission. However, this can suffer from dust obscuration of the emission lines, and emission-filling of H Balmer lines, and it requires a high signal-to-noise ($\ge$ 10 per pixel). In addition, there has been some debate over the criteria for selecting \ea\ galaxies, with some authors choosing \ha , H$\delta$ and [OII], with differing line strength criteria (e.g. Goto 2004, 2005), and some including H$\gamma$ as well as \ha, H$\delta$ and [OII] (e.g. Zabludoff et al. 1996; Blake et al. 2004). 

Here, we use a purely data-driven rectified factor analysis data model, developed in Nolan et al. (2006), on the spectra of early-type galaxies in the SDSS Data Release 4 (DR4). This was specifically designed to exploit large data sets in a model-independent way. It uses all the data present in the spectra, i.e. the spectral features and the overall shape of the continuum, to quantify the relative strength of recent star formation in early-type galaxies extracted from the database, in a rapid, objective and robust manner. Our method allows us to reconstruct even partial spectra, without the need to measure individual lines accurately. The advent of large surveys, such as the Sloan Digital Sky Survey allows us to study large samples of these rare objects ($<$ 1 $\%$ of the overall zero-redshift galaxy population, Zabludoff et al., 1996), in a way that has not been previously possible.

It has recently been shown that \ea\ galaxies occur more frequently at high redshift (1 $\leq$ z $\leq$ 2) compared with the local universe, at least amongst the massive red galaxy population (Doherty et al. 2005, Le Borgne et al. 2006). They may, therefore, represent a critical phase in the evolution of early-type galaxies. However, it is difficult to constrain the properties of these objects at high redshift, due to the difficulty of measuring absorption lines accurately at that redshift, and in obtaining statistically significant spectroscopic samples. Studying large samples at low redshift is therefore vital for the understanding of these objects.

In this work, we use our data modelling to identify early-type galaxies (ETGs) which have a young stellar population, suggesting that they have undergone a recent burst of star formation. We compare these galaxies with \ea\ galaxies, selected in a completely independent manner, and investigate the environment of our galaxy samples, in order to understand the physical mechanisms leading to the cessation of star formation.

In Section \ref{sample}, we define the selection of the early-type sample, and the \ea\ subsample for comparison. We discuss the factor analysis of the early-type spectra that is used to reveal young stellar populations in Section \ref{model}. In Section \ref{anal}, we compare our results from the analysis of early-type galaxies with the sample of \ea\ galaxies. In  Section \ref{environ}, we investigate the relationship between the
galaxies in our various samples and their local environment. Our conclusions are discussed in Section \ref{conc}.

We use $H_{0} = 70$ kms$^{-1}$Mpc$^{-1}$, $\Omega_{m} =$0.3 and $\Omega_{\Lambda} =$0.7 throughout.

\section{The sample}\label{sample}

Our sample of early-type galaxy spectra was selected from the SDSS~DR4 spectroscopic catalogue. We follow the criteria of Bernardi et al. (2003) to select early-type galaxies:

\begin{itemize}

   \item the likelihood that a de Vaucouleurs profile fits the radial surface brightness distribution is at least 1.03 times the likelihood that the radial profile is exponential;
   \item the concentration index $r_{90} / r_{50} > 2.5$ in the $i$-band;
   \item the $r$-band Petrosian magnitude $<$ 17.5
   \item the photometric signal-to-noise $>$ 4
   \item the galaxy is not blended, de-blended or a child galaxy
   \item a surface brightness profile has been fitted, and it is not saturated
   \item redshift, $z < 0.3$ and the confidence in the redshift, {\it zConf}$\!>\! 0.95$.
\end{itemize}

We also select \ea\ galaxies from the SDSS~DR4. Here, we use the criteria of Goto (2005), so that we can compare the results of our factor analysis of the early-type galaxy sample with their sample of \ea s. The criteria are:

\begin{itemize}
  
   \item \ha\ equivalent width (EW) $<$ 3.0 \AA
   \item H$\delta$ EW $<$ - 5.0 \AA
   \item \oii\ EW $<$ 2.5 \AA
   \item S/N $>$ 10  
   \item z $>$ 0.01,
  
\end{itemize}

\noindent
where emission lines have positive values. The sample is not restricted by any morphological criteria.

This gives us a sample of over 26,000 early-type galaxy spectra, and 765 Goto \ea s, which we de-redshift and re-bin at the SDSS full-width half-maximum (FWHM) spectral resolution, with the observational errors supplied re-binned correspondingly in quadrature.

In this work, we consider only those galaxies in our samples which lie in the range 0.06 $<$ z $<$ 0.14. Around 50\% of the galaxies in our ETG sample lie within this range, and restricting ourselves to a limited redshift range minimises any effects arising from the fixed size of the SDSS fibre diameters. At z $=$ 0.1, the central $\sim$ 5.5 kpc are covered, which represents 20$-$40\% of the total stellar light, which should be a good representation of the bulk stellar light. The emission lines used to select \ea\ galaxy spectra are measured from the same spectra that we decompose using our data model, so we are of course comparing results extracted from exactly the same region of each galaxy.

The redshift cut means that, in this work, we do not explore the evolution of ETGs with young stellar populations with redshift. However, we also need not worry about $k$-corrections to the magnitudes. The redshift range 0.06 $<$ z $<$ 0.14 is less than 1 Gyr of stellar evolution; $k$-corrections would probably introduce uncertainties at least on this scale. Hence, we use a magnitude limit of $r <$ 17.5 for all our samples, where $r$ is the Petrosian $r$-band magnitude. The SDSS ~DR4 spectroscopic catalogue is complete to this limit. 

\section{Modelling the spectra}\label{model}

In order to analyse the spectra of our early-type galaxy sample, we
model the data using the rectified factor analysis data model that we
developed in Nolan et al. (2006). This model-independent technique is
able to recover physically meaningful components of the observed
spectra in a timely manner. Our approach and the underlying rationale
for choosing the rectified factor model for modelling the data are
described in detail in Nolan et al. (2006, \S 4.1), along with the details of
an implementation employing variational Bayesian techniques (Nolan et al. 2006, \S3.1). This
modelling is an unsupervised technique, i.e. a data-driven exploratory
analysis tool, designed to help us discover the underlying structure
of the data and to derive features for either interpretation or
subsequent quantitive analyses. Unsupervised techniques are machine learning approaches, which do not require {\it a-priori} specified targets for the training examples but instead, the data density is modelled. This model also includes latent variables that are then used to infer patterns in the data that are hoped
to reveal useful relationships.

\begin{figure}

\includegraphics[width=6.0cm,angle=-90]{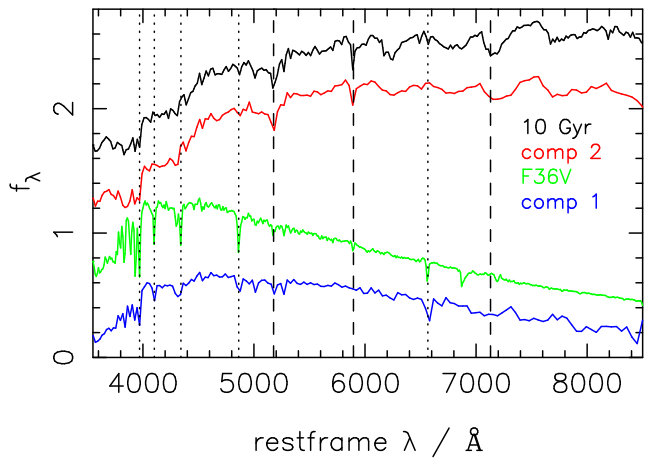}

\caption{\small{ From {\bf top to bottom}: a 10 Gyr, 2.5 \Zsolar\ single stellar population model (Bruzual \& Charlot 2003, {\bf black}); component 2 from our analysis ({\bf red}); a super-solar metallicity F star spectrum (Santos et al. 1995 {\bf green});  component 1 from our analysis ({\bf blue}). The dotted lines mark some of the absorption features in the spectra which are
typically strong in young stellar populations, and the dashed lines
mark some of the absorption features which are typically strong in
old, metal-rich stellar populations. From left to right, the
absorption line species are: H$\varepsilon$ (3970
\AA), H$\delta$ (4102\AA), H$\gamma$ (4340\AA), H$\beta$ (4861
\AA), Mgb (5175\AA), NaD (5893\AA), H$\alpha$, TiO
(7126\AA). It can clearly be seen that the first component ({\bf blue}) represents a younger stellar contribution ({\bf green}), whereas the second component ({\bf red}) represents a mature stellar population ({\bf black}).}}\label{spectra}

\end{figure}

The model is first trained on half of the SDSS early-type galaxy
spectra ($\sim$ 13000 galaxies), and the recovered components are then
used to reconstruct the remaining ETG spectra, and the spectra in the
Goto \ea\ sample (G\ea). Following Nolan et al. (2006, see \S 4 and ~Fig. 4), we chose to
have two spectral components, which was shown to be sufficent to
represent the bulk of the stars in early-type
galaxies. Fig.~\ref{spectra} shows the two component spectra recovered
from the factor analysis, together with a 10 Gyr, 2.5 \Zsolar\ single
stellar population model (Bruzual \& Charlot 2003) and a typical
super-solar F star spectrum (Santos et al. 1995) for
comparison. Features typical of a mature, metal-rich stellar
population, and a young stellar population are marked on the plot. 
It can be seen that the first component represents a contribution dominated
by younger stars ($\sim$1--4~Gyr for a typical F star at the main
sequence turn-off point). It has strong hydrogen Balmer absorption
lines and a corresponding lack of evolved metal absorption
lines. The second component has the characteristic shape and
spectral features of a mature, metal-rich stellar population, such as
those that are expected to dominate in early-type galaxies. This
component has a well-developed 4000\AA\ break, typical of such
populations. Clearly, the two components recovered represent young and old
stellar contributions to the spectra of ETGs, although of course they
cannot be exactly interpreted as single stellar populations. Although the second component is well-fit by the 10 Gyr stellar population, the first component is slightly redder than that of the F star. This probably reflects both a more complex star formation history than a simple two-component model, and that the 'real' spectrum is that of the integrated young stellar population, rather than that of a single stellar type. Ferreras et al. (2006), using a principal components analysis on the
3500$-$7500\AA\ spectra of 30 ellipticals in groups and the field,
recover similar representations of old and young stellar populations
in their first two components. Their results are also consistent with
the representation of the bulk of stars in early-type galaxies by two
components.

\section{Comparison of the early-type galaxies and the \ea\ sample}\label{anal}

Within our chosen redshift range, we are left with 13,111 ETGs. 
Keeping only those galaxies in the G\ea\ sample which have
\ha, H$\beta$ and [OII] securely determined leaves us with 196 G\ea
s. Within the ETG sample, we define a sub-sample of ETGs with young
stellar populations (YSPETs). These are galaxies for which a1, the
contribution (or weight) of the `young' data model component is
greater than a2, the weight of the `old' data model component. By
comparison with stellar population models, we estimate a1 $>$ a2 is
equivalent to an approximate lower limit to the young stellar
population of \gs 10\% by stellar mass. YSPETs account for 17\% of the
ETGS (2245 galaxies).

\begin{figure}

\includegraphics[width=6.2cm,angle=-90]{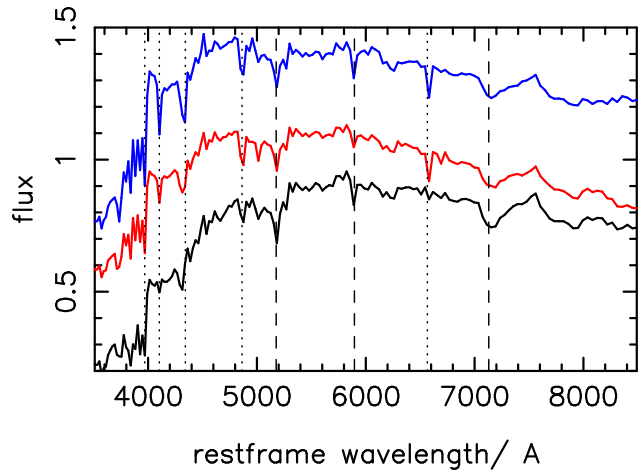}

\caption{\small{Mean spectra for the whole ETG sample ({\bf black, bottom}), the YSPET sub-sample ({\bf red, middle)}), and the G\ea\ sample ({\bf blue, top}). The spectra are normalised to unity at 5500 \ang\ and the vertical normalisation adjusted for comparison. Longwards of H$\delta$, the \ea\ and the YSPET populations look very similar. The mean ETG spectrum is flatter, and does not have such pronounced hydrogen Balmer absorption lines. The dotted and dashed lines are the same as for Fig.~\ref{spectra}.   } }\label{meanspec}

\end{figure}

\begin{figure}

\includegraphics[width=5.0cm,angle=-90]{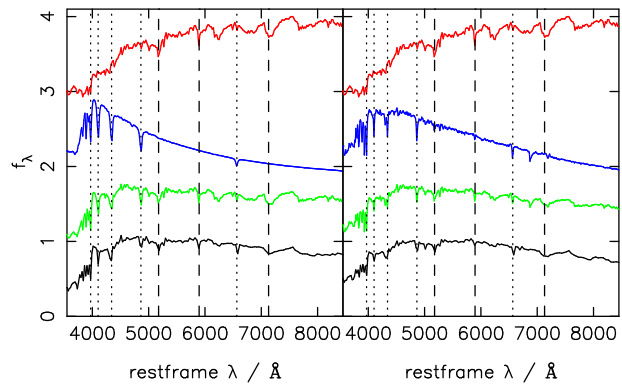}

\caption{\small{ {\bf LHS, from top to bottom:} A 10 Gyr, 2.5 \Zsolar\ stellar population model (Bruzual \& Charlot 2003, {\bf red:}); An ``A star'' spectrum (Kurucz 1993, {\bf blue}); an \ea\ spectrum, which is a linear combination of the first two spectra ({\bf green}); and the mean G\ea\ spectrum ({\bf black}). {\bf RHS:} as for the LHS, but here we use a super-solar metallicity ``F star'' (Santos et al. 1995, {\bf blue}) instead of the A star, and the lowest ({\bf black}) spectrum is the mean spectrum of the YSPET population. The mean G\ea\ spectrum, as one would expect, is very similar to our constructed \ea\ spectrum. The mean YSPET spectrum, however, looks more like a `E+F' spectrum. The dotted and dashed lines are as for Fig.~\ref{spectra}.}}\label{eaplot}

\end{figure}

In Fig.~\ref{meanspec} we plot the mean spectra for each of these three catagories, with the vertical normalisation adjusted for ease of comparison. The mean G\ea\ and YSPET spectra are very similar longwards of H$\delta$, and both have stronger hydrogen Balmer absorption lines and a steeper 5000$-$8000\AA\ slope than the ETG spectrum. This is characteristic of the presence of contributions to the flux from young stars. The G\ea\ mean spectrum shows contributions from the characteristic features of even younger stars than does the YSPET mean spectrum. It has stronger hydrogen Balmer absorption lines, enhanced flux just longwards of 4000\AA, and the flux falls off more steeply towards the blue end  of the spectrum shortwards of 4000\AA\ (see Fig.~\ref{eaplot}).

In Fig.~\ref{eaplot} we show the likely stellar composition of the G\ea\ and YSPET galaxies. The \ea\ galaxies, as one would expect, resemble the linear combination of a mature, metal-rich stellar population (the `E' component) together with a population dominated by young ($\sim$ 0.7 Gyr) A stars. The mean YSPET  spectrum is similar, but in this case, the younger stellar component is better-represented by F stars, which dominate the flux of a young stellar population at a later epoch than A stars ($\sim$ 1$-$ 4 Gyr). Hence, our factor analysis has allowed us to define a sample of early-type galaxies with young-intermediate aged stars, which, as F stars dominate the integrated spectrum of a stellar population for longer than the short-lived A stars, presents us with a larger sample of galaxies which have undergone recent star-forming activity than \ea\ samples are able to.

\begin{figure*}
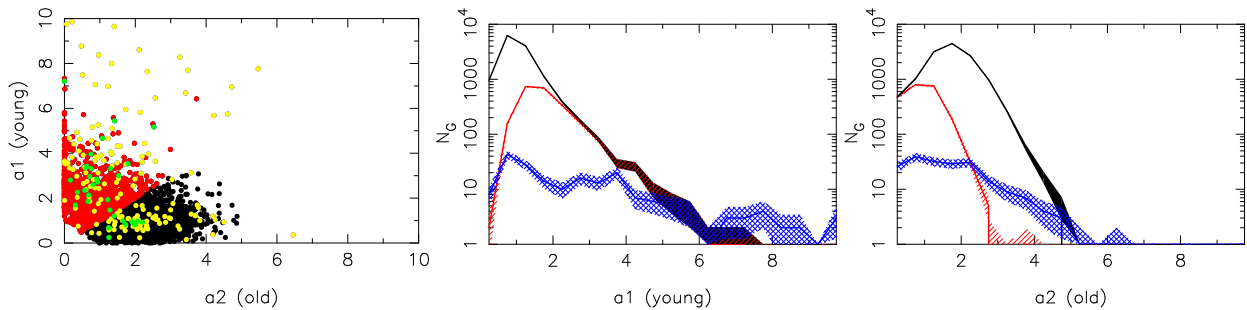


\includegraphics[width=4.0cm,angle=-90]{fig4a.epsf}
\includegraphics[width=4.0cm,angle=-90]{fig4b.epsf}
\includegraphics[width=4.0cm,angle=-90]{fig4c.epsf}

\caption{\small{Distribution of the weights a1 (the weight of the `young' component from our data modelling) and a2 (the weight of the `old' component from our data modelling), for: all the early-type galaxies in the sample ({\bf black}); Goto \ea s ({\bf blue, cross-hatched }); YSPETs ({\bf red, hatched}). In the {\bf left} plot, the G\ea\ sample is further divided into those that would be classified in our scheme as early-types ({\bf green}) and those that would not ({\bf yellow}). }}\label{na1a2}

\end{figure*}

Furthermore, in Fig.~\ref{na1a2}, we plot the number densities of the
weights of the data model components for our three classes of
spectra. We have further divided the G\ea\ galaxies into those that
would be classified in our scheme as early-types, and those that would
not. Only $\sim$ 17 \% of the G\ea s (34) are typical early-type
galaxies, of which 2/3 are in the a1 $>$ a2 region. The distribution
of the G\ea s is less peaked in either a1 or a2 than the YSPETs, which
broadly follow the same distribution as the ETGs.

That most of the G\ea\ galaxies are not ETGs is consistent with the idea that they are the result of galaxy-galaxy mergers or interactions, and are at an earlier epoch in their post-merger evolution than the YSPETs. Their young stellar component is consistent with being less than 1 Gyr old. This is less than the typical time for the colours and surface brightness distributions of merger remnants to relax into those which are typical of ETGs, and which we use to identify ETGs in the SDSS catalogue. Of course, it should also be noted that if G\ea s are quenched spirals, whose star formation has been recently halted by ram-pressure or tidal stripping, one would also expect them not be classified as ETGs under our scheme.

\section{The environmental dependence}\label{environ}

In order to investigate the mechanisms via which \ea s are formed, we
have quantified the environment of our sample galaxies. First, we find
$\sigma_{10}$ and $\sigma_{5}$, the projected distance between the
galaxy and its 10th and 5th nearest neighbours in the SDSS
spectroscopic catalogue. We use a Petrosian magnitude limit, $r <$
17.5, and do not $k$-correct our magnitudes, due to the limited
redshift range of our sample (see \S \ref{sample}). The SDSS~DR4
spectroscopic survey is $\sim$ 92 $\%$ complete (Ani Thakar, private
communication), so we are able to take advantage of the spectroscopic
redshifts in determining environment.

We cut in peculiar velocity at $\pm$ 1000 kms$^{-1}$ from each galaxy, to attempt to avoid background or foreground objects, but include cluster members if the target galaxy belongs to a cluster. We have not attempted here to determine cluster membership, and a more sophisticated analysis is reserved for a later paper. Here, we are only interested in the projected local galaxy density.

The projected distances are then transformed into an estimate of the local projected surface density, $\Sigma_{10}$ and $\Sigma_{5}$. We have made a cut in $\sigma_{n}$ to allow for the effects of the survey boundary, which could lead to an underestimate of local density for those galaxies in low-density regions. Following Clemens et al. (2006) and Miller et al. (2003) we cut at $\sigma_{n}(max) =$ 7.2 Mpc. 

\begin{figure*}
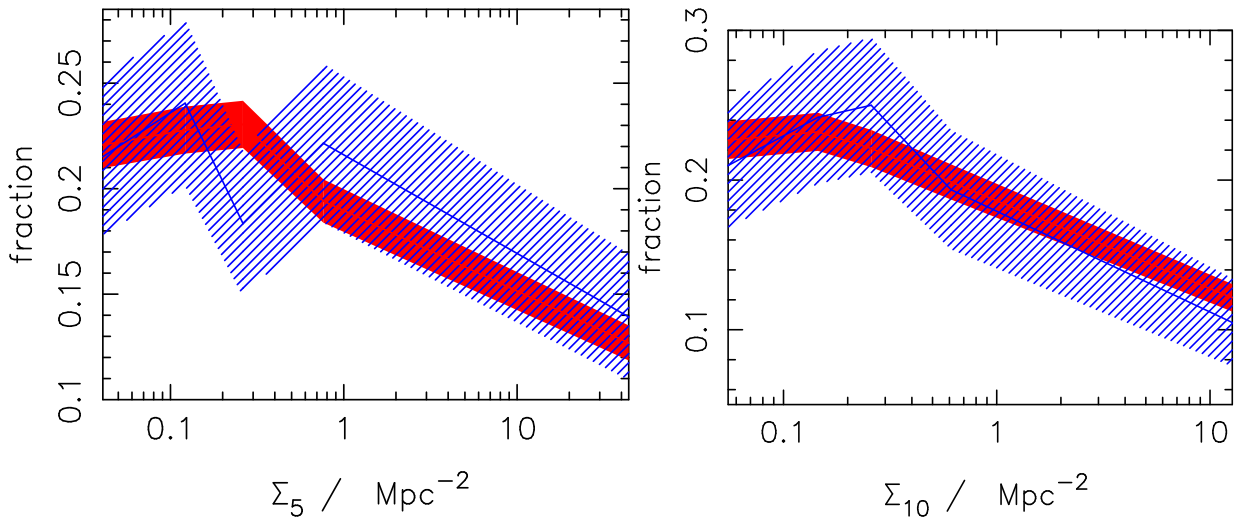

\includegraphics[width=6.8cm,angle=-90]{fig5a.epsf}
\includegraphics[width=6.8cm,angle=-90]{fig5b.epsf}
\caption{\small{ Normalised distributions of local surface density: early-types with a2 $>$ a1 ({\bf red, solid}) and Goto's \ea s ({\bf blue, cross-hatched}) and $\Sigma_{5}$ ({\bf top}), $\Sigma_{10}$ ({\bf bottom}). The populations are normalised so that the area under each line represents 100$\%$ of the galaxies in that sample.}}\label{ndens}
\end{figure*}

In Fig.~\ref{ndens}, the normalised number distributions of local galaxy surface density, $\Sigma_{10}$ and $\Sigma_{10}$, are shown for the YSPET and the G\ea\  galaxy populations. The similarities between the samples can be clearly seen. The distribution of the G\ea s follows that of the YSPETs in both cases. The fraction of G\ea s perhaps peaks at around $\Sigma_{10} \sim $ 0.1 $-$ 0.3 Mpc$^{-2}$, and both populations fall off strongly towards higher density environments. The fraction of YSPETs flattens off after this peak in $\Sigma_{10}$, and in $\Sigma_{5}$, the YSPET distribution is consistent with a flattening off at surface densities less than 0.2$-$0.3. 

Performing a Kolmogorov-Smirnov test on the G\ea\ and YSPET
populations returns the probabilities, ($p - 1) = $ 4 $\times$ 10$^{-4}$
($\Sigma_{5}$) and 0.3 ($\Sigma_{10}$) that the two samples are taken
from different distributions. Therefore, in both cases, despite our
completely independent selection procedures, the distributions of the
two populations are not significantly different, consistent with the
hypothesis that our YSPETs are in fact evolved \ea\ galaxies (`E+F'
galaxies). The decreasing fraction of \ea s and YSPETs towards
increasing densities strongly argues for the hypothesis that the majority of G\ea s or
YSPETs are {\it not} the product of cluster-related processes
(e.g. ram-pressure stripping). This confirms the result of Goto (2005)
that \ea\ galaxies are likely to be the product of galaxy-galaxy
mergers and interactions in a group environment.

Studies of star-forming galaxies of all morphologies in the SDSS and
the Two Degree Field Galaxy Redshift Survey (2dFGRS) (e.g. Lewis et
al. 2002; Gomez et al. 2003; Balogh et al. 2004; Poggianti et
al. 2006) find that the star formation rate falls off steeply with
increasing density, but converges to the field galaxy distribution at
a projected density of $\sim$ 1 Mpc$^{-2}$. In addition, Balogh et
al. (2004) find that the fraction of star-forming galaxies of all
morphologies falls steeply with respect to $\Sigma_{5}$ \gs 0.4
Mpc$^{-2}$, flattening off at densities less than this, in good
agreement with our results (see their Figs~5 and 6).

The peak in fraction at $\Sigma_{10} \sim $ 0.1 $-$ 0.3 Mpc$^{-2}$ may
indicate the optimum local galaxy surface density for galaxy-galaxy
mergers and interactions, representing a trade-off between sufficient
density for frequent galaxy-galaxy encounters, and slow enough
relative velocities that those encounters lead to an
interaction. Investigation of the results of N-body simulations would
be able confirm or reject this. However, there is support for this
argument in the recent results of Ferreras et al. (2006). They find a
higher mass fraction of younger stars in some group galaxies than in
the field galaxies, implying a more complex star formation history for
galaxies in groups than in the field, in qualitative agreement with
our tentative peak of recent starburst activity.

Further evidence for a peak in star-formation with respect to local galaxy density is found by Porter and Raychaudhury (2006, in preparation). Along filaments joining rich clusters, they find a peak in star formation at 3$-$4 Mpc from cluster centres. However, a careful assessment of group, cluster and filament membership is necessary, as the precise location of the peak depends upon whether a galaxy is a member of any these environmental classes.

Many authors have found that the star-formation rate begins to fall at densities \gs 1 Mpc$^{-2}$ (e.g. Lewis et al. 2002; Gomez et al. 2004; Balogh et al. 2004; Poggianti et al. 2006; Porter and Raychaudhury 2006, in preparation). However we find, as do Balogh et al. (2004), that the fraction of star-forming galaxies starts to fall off at lower densities (0.1$-$0.4 Mpc$^{-2}$), with the caveat that peaks and breaks with respect to density depend on a measurement of environment more complex than simple projected local galaxy density, as mentioned above. This is suggestive that actively star-forming disk galaxies have their (moderate compared with disc-disc merging starbursts) star-formation quenched in group environments, decreasing the active-to-passive galaxy ratio. At the same time, more disc-disc merging or interaction occurs in the group environment, with associated massive star formation, keeping the star-formation rate high at higher densities, whilst contributing to the fall in the number of disc galaxies. Hence, both morphological and spectroscopic data are required in order to study these different processes effectively.

\section{Conclusions}\label{conc}

We have successfully used our data-driven rectified factor analysis to
identify early-type galaxies with young stellar populations, without
having to apply the far more computationally expensive method of
fitting detailed stellar population synthesis models to a sample of $>$ 13,000 ETGs in the redshift range 0.06 $<$ z $<$ 0.14. We have used the
sample generated as a result
to map the
relationship between early-type galaxies, \ea\ galaxies and early-type
galaxies with young stellar populations. Our analysis recovers two
components from the set of early-type galaxy spectra, one of which
represents the contribution to the flux from mature, metal-rich stars,
and the other which represents the contribution from F stars, $\sim$1--4~Gyr 
old. Our method allows us to reconstruct spectra even when
the observed spectra are incomplete. We do not need high
signal-to-noise measurements of individual lines. We identify those
early-types which have undergone recent ($\sim$1--4~Gyr) star
formation, with the young stars accounting for at least 10\% of their
stellar mass (YSPETs). At redshifts 0.06 $<$ z $<$ 0.14, $\sim$17\% 
 of ETGs are classified as YSPETs. We propose that these YSPETs
are `E$+$F' galaxies, whose spectra are a linear combination of a
typical early-type galaxy spectrum and that of an F star, analogous to
\ea\ spectra. 

We also select a sample of \ea\ galaxies in a completely independent
way, without using  the morphological criteria we used 
to identify early-type galaxies, but instead relying on 
the emission and absorption-line criteria of Goto (2005). The
characteristic `\ea' spectra of the Goto-selected sample, and the fact
that many of these \ea s do not fulfil the colour and concentration
criteria for typical early-type galaxies, suggests that these galaxies
are in an early ($<1$ Gyr) post galaxy-galaxy merger or interaction
phase. Goto (2005) notes that most of the \ea s in his sample have
concentrations as high as typical ellipticals ($C_{in}\! <\! 0.4$), so it
is likely that many of these galaxies are not included in our
early-type sample because they are too blue, indicating that massive
(A star and earlier) stars are still burning. If \ea s are
recently quenched star-forming galaxies, we would expect them to be
located preferentially in dense cluster regions, whereas we see the
majority residing in field or group environments, and the number falls
off towards denser environments, in agreement with Goto's
result. Their high concentrations and preferentially low-density
environments argue strongly that these are the products of
galaxy-galaxy mergers or interactions, which will eventually settle
down to become typical early-type galaxies.

We note that there is in fact a slight peak in the local galaxy
surface density distribution of \ea\ galaxies at 0.1--0.3
Mpc$^{-2}$. This may represent the optimum local galaxy surface
density for galaxy-galaxy mergers and interactions. Realistic cosmological simulations
should be able to confirm or reject this hypothesis.

Given that the distribution of local galaxy surface
density of the G\ea\ galaxies and that of the independently determined YSPET
populations are not significantly different, one might expect a similar history for both sets of
galaxies. We hypothesise that the YSPETs, with typical ETG colours and
concentrations, and `E$+$F' spectra, represent an intermediate
evolutionary stage between \ea s and early-types. As this is a
longer-lived stage than \ea\ galaxies (F stars persist on the main sequence
for longer than A stars, and dominate the integrated flux at ages \ls
5 Gyr), we are able to identify a larger sample of these galaxies than
is possible with the very rare \ea s. We therefore have identified a
powerful tool for studying in detail, both globally and individually, the
evolution of early-type galaxies.

Our future work will compare the dependence on local density of \ea\ and E$+$F galaxies that we have recovered here, with those recovered from simulations. We intend to perform the same analysis on 2dFGRS galaxies and a sample of high-redshift galaxies for comparison. We also intend to investigate the relationship between \ea\ and E$+$F galaxies by studying their kinematics in detail using integral field spectroscopy.

\section{Acknowledgements}
Thanks to Ani Thakar at the SDSS helpdesk for valuable assistance with
the SDSS archive.  Funding for the SDSS and SDSS-II has been provided
by the Alfred P. Sloan Foundation, the Participating Institutions, the
National Science Foundation, the U.S. Department of Energy, the
National Aeronautics and Space Administration, the Japanese
Monbukagakusho, the Max Planck Society, and the Higher Education
Funding Council for England. The SDSS Web Site is
http://www.sdss.org/.

\end{document}